\documentclass[prb,longbibliography,twocolumn,nopacs,preprintnumbers,notitlepage,amsmath,amssymb,superscriptaddress]{revtex4-2}

\usepackage{epsfig}
\usepackage[colorlinks=true,urlcolor=blue,citecolor=blue,linkcolor=blue]{hyperref}
\hypersetup{breaklinks=true}
\usepackage{color}

\usepackage{dcolumn}   
\usepackage{bm}        
\usepackage{amsmath}
\usepackage{amsfonts}
\usepackage{bbm}
\usepackage{float}
\usepackage{enumerate}
\usepackage{tikz}
\usepgflibrary{arrows.meta}
\usepackage{physics}
\usepackage{mathrsfs}
\usepackage[normalem]{ulem}
\usepackage{graphicx}  
\usepackage{subfigure}
\usepackage{makecell}
\usepackage{lipsum}
\usepackage{romannum}

\begin{document}
\title{Large critical fields in superconducting Ti$_{4}$Ir$_2$O from spin-orbit coupling}
\author{Hao Wu}
\author{Tatsuya Shishidou} 
\author{Michael Weinert} 
\author{Daniel F. Agterberg}
\affiliation{Department of Physics, University of Wisconsin--Milwaukee, Milwaukee, Wisconsin 53201, USA}
\date{\today}

\begin{abstract}
The recently synthesized $\eta$-carbide-type superconductors exhibit large critical fields. A notable example is Ti$_4$Ir$_2$O, for which the upper critical field strongly violates the Pauli paramagnetic limit, behavior that is unusual for cubic materials that preserve inversion symmetry. Here, by combining density functional theory (DFT) and analytic modeling, we provide an explanation for this enhanced Pauli limiting field. We show that the nonsymmorphic Fd$\overline{3}$m symmetry implies that the electronic states near the X points exhibit strong spin-orbit coupling (SOC), which leads to a vanishing effective $g$-factor and enables the enhanced Pauli limiting field. Furthermore, our DFT results reveal a Van Hove singularity (VHS) peak near the X points, accounting for $\sim$65\% of the total density of states (DOS), occurring near the chemical potential. We propose that the strong SOC and enhanced DOS in the vicinity of the X points provide the origin of the observed enhanced critical field. This leads to a prediction that the magnetic field will lead to a strongly momentum-dependent gap suppression. The gap due to electronic states away from (near to) the X points will be rapidly (slowly) suppressed by fields.
\end{abstract}

\maketitle

{\it Introduction.} The investigation of large upper critical fields in superconductors is of technological importance since it holds potential for significant advances in high-field magnetic applications. High upper critical fields enable superconductivity for magnetic resonance imaging (MRI), nuclear magnetic resonance (NMR) spectroscopy, mass spectrometers, magnetic separation processes, high-speed magnetic levitation (Maglev) trains, high-energy particle accelerators, and magnetic confinement fusion reactors. Superconductors such as NbTi and the A15 superconductor Nb$_{3}$Sn are widely used in these high-field applications due to their high upper critical fields~\cite{okuno2004superconducting, stewart2015superconductivity, banno2023low}. Hence it is important to understand when such high fields occur. In this context, the large critical fields in $\eta$-carbide-type superconductors offer a promising avenue of exploration.

The $\eta$-carbide-type superconductors crystallize in the cubic space group Fd$\overline{3}$m(No.\ 227)~\cite{taylor1952new, kuo1953formation, mackay1994new}. It has been found that Zr$_4$Rh$_2$O$_{0.7}$~\cite{ma2019superconductivity}, Zr$_4$Rh$_2$O~\cite{ma2019superconductivity,watanabe2023observation}, and Ti$_4$Rh$_2$O~\cite{ma2021group}, exhibit upper critical fields close to the Pauli paramagnetic limit~\cite{chandrasekhar1962note, clogston1962upper, sarma1963influence, maki1964pauli}, while Nb$_4$Rh$_2$C$_{1-\delta}$~\cite{ma2021superconductivity}, Ti$_4$Co$_2$O~\cite{ma2021group}, Ti$_4$Ir$_2$O~\cite{ma2021group, ruan2022superconductivity}, and Zr$_4$Pd$_2$O~\cite{watanabe2023observation}, exhibit large upper critical fields that strongly violate the Pauli limit. Understanding the mechanisms of these large critical fields is the goal of this work.

Insight into the origin of the enhanced Pauli field comes from the observation that a spin-orbit coupling (SOC)-driven avoided band crossing appears near the chemical potential~\cite{shi2023pressure}. Here, we build upon this insight and show that in the vicinity of the X points there is an enhanced density of states (DOS), and that symmetry requires a large SOC that drives an unusual Bloch spin structure -- anomalous pseudospin -- and it is this feature that enables the large observed critical fields. 

Specifically, we have carried out density functional theory (DFT) calculations on Ti$_{4}$Ir$_2$O. We find that,  centered at the X points, there are topological changes in the Fermi surface, Lifshitz transitions, which are accompanied by a Van Hove singularity (VHS) in the DOS~\cite{lifshitz1960anomalies,blanter1994theory}. In the vicinity of the VHS, the X-centered Fermi surfaces dominate the total DOS ($\sim$65\%), and this VHS lies just below the chemical potential, suggesting that these X-centered Fermi surfaces play an important role in the superconducting properties. 

\begin{figure*}   
\centering
\includegraphics[width = 1\textwidth]{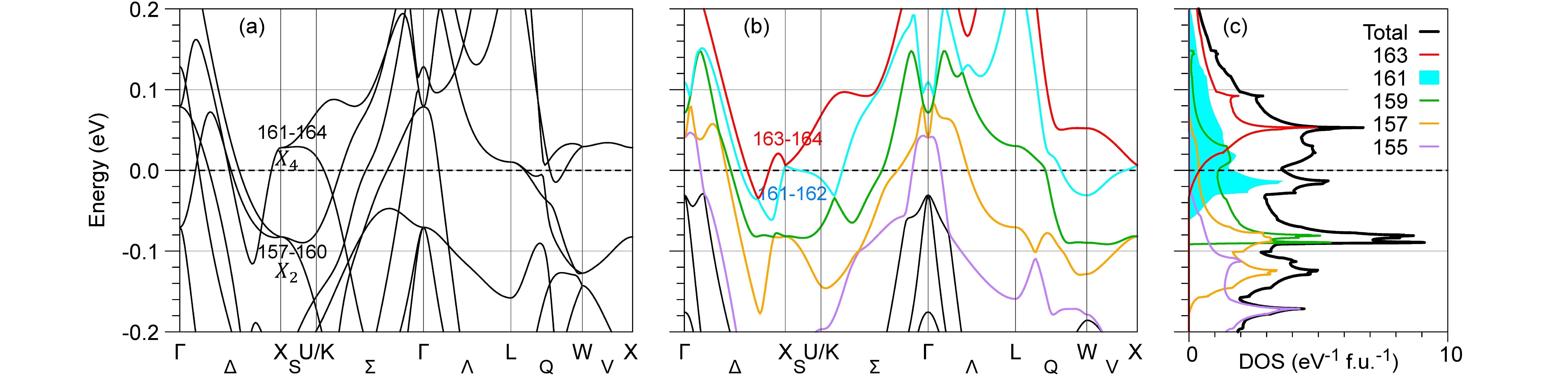}
\caption{Energy band dispersion near the chemical potential $E=0$. 
 (a) without and (b) with spin-orbit coupling. (c) Band-resolved density of states with spin-orbit coupling considered.}
\label{band}
\end{figure*}
\begin{figure*}   
\centering
\includegraphics[width=1\textwidth]{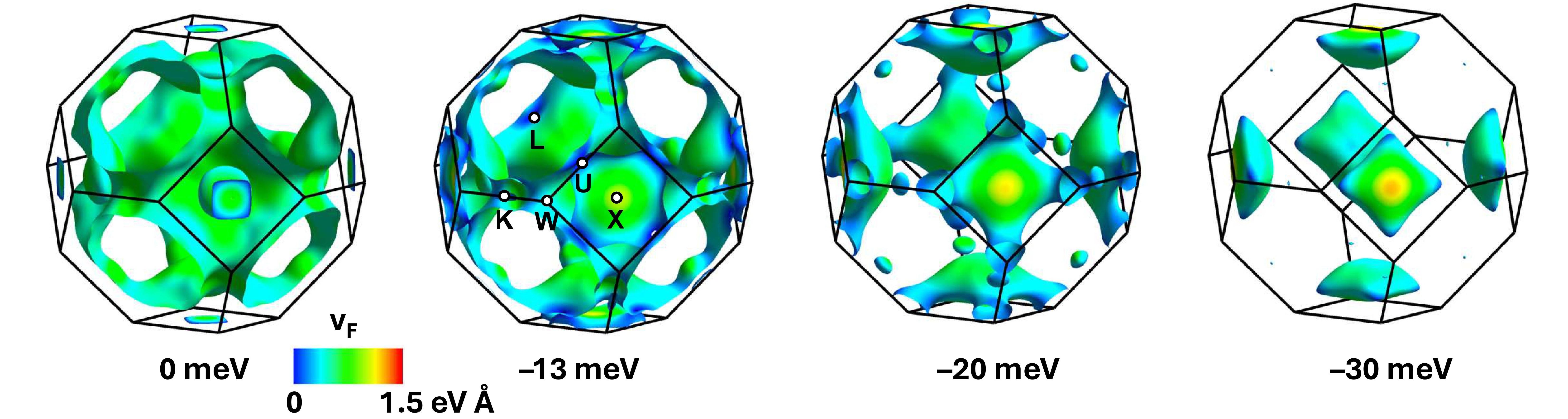}
\caption{Fermi surfaces of bands 161-164, illustrating the distinct topological changes as the chemical potential is lowered.}
\label{FS}
\end{figure*}

As mentioned above, an additional important feature of our theory is that, near the X points, SOC is strong and leads to an anomalous Bloch pseudospin that only couples to magnetic fields oriented in a single direction.
In particular, each X point 
lies at the intersection of two orthogonal momentum lines on which a four-fold degeneracy is guaranteed by symmetry when SOC is not included. When SOC is included this four-fold degeneracy splits into two two-fold degenerate Kramers pseudospin pairs. These pseudospin pairs have the feature that they do not couple to magnetic fields orthogonal to the momentum lines along which the original four-fold degeneracy appears. This lack of coupling ensures enhanced Pauli limiting fields in the superconducting state.  Our analysis of these momentum lines generalizes our earlier analysis of a similar anomalous pseudospin that appears on momentum planes \cite{suh2023superconductivity}.

{\it DFT calculations.}
Our DFT calculations are carried out using the full-potential linearized augmented plane wave (FLAPW) method~\cite{weinert2009flapw}. The generalized gradient approximation~\cite{perdew1996generalized} for the exchange-correlation, wave function and potential energy cutoffs of 16 and 200 Ry, respectively, muffin-tin sphere radii of 1.25 \r{A} for Ti and Ir and 0.8 \r{A} for O atoms, $8^3$ k-point mesh for self-consistent field calculation, and $100^3$ k-point mesh for density of states and Fermi surface calculations are employed. The Ti 3$p$ and Ir 5$p$ orbitals are treated as core states. Relativistic effects are fully taken into account. The structural parameters determined experimentally at ambient pressure~\cite{ruan2022superconductivity} are used.

Figures~\ref{band} (a) and (b) show the energy band dispersion calculated along high symmetry lines, which is in fair agreement with previous calculations~\cite{ma2021group}. At the X point (0,1,0), the bands are four-fold degenerate with and without SOC. Along a zone-boundary line V (u,1,0), connecting X to W, the four-fold degeneracy is lifted only through SOC, which plays an important role in our calculations of the Pauli limiting field later. Along the $\Delta$ line $(0,v,0)$ connecting from X to $\Gamma$, the band character at X implies different behavior for the non-SOC band splitting: Vanishing for bands belonging to the $X_4$ irreducible representation (IR), as defined in \cite{aroyo2011crystallography,aroyo2006bilbao,aroyo2006bilbao2,elcoro2017double}, and non-vanishing for the $X_2$ IR at the X point. The bands 161-164 belong to the $X_4$ IR: their splitting on the $\Delta$ line arises solely from SOC. In addition, they show anisotropic band curvature, with negative and positive curvatures along the $\Delta$ and V lines, respectively.  Furthermore, the $k$-linear SOC splitting also exhibits highly anisotropic behavior (about five times larger in the $\Delta$ direction than in the V direction). 

As demonstrated in Fig.~\ref{band}(c), the SOC-split lower-energy band 161-162 shows a pronounced peak in the DOS at energies ($\sim -13$ meV) very close to the theoretical chemical potential, dominating the total DOS ($\sim$65\%). This proximity to the chemical potential suggests that, depending upon the electron/hole doping (e.g., through vacancies/impurities inherent in the samples) or applied pressure, the X-centered bands 161-164 may move closer to the chemical potential and become the most important bands for superconductivity. Fig.~\ref{FS} shows the Fermi surfaces from the bands 161-164. The band 161-162 provides two elements: a small square sheet attached to each X face and a large element scattering inside the Brillouin zone. If the chemical potential is lowered, the square sheet grows, and eventually the two elements merge to form six separate elements located slightly away from the X faces. This continuous change of the Fermi surface, which involves a change in topology, accompanies the enhanced DOS peak discussed above. 

{\it $kp$ theory analysis.} Our DFT results motivate an analysis of a symmetry-based $kp$-like Hamiltonian valid near the X points. The DFT calculations reveal that the $X_4$ IR provides the relevant electronic states near the chemical potential. The electronic $X_4$ IR is four-fold degenerate with a two-fold sublattice degeneracy and a two-fold spin-degeneracy. We use Pauli matrices $\tau_i$ ($\sigma_i$) to describe the sublattice (spin) degrees of freedom. Following the approach developed in Ref.~\cite{suh2023superconductivity}, we find   
\begin{equation}
\begin{gathered}
H(\mathbf{k})=\varepsilon_{0, \mathbf{k}}+t_1 k_x k_z \tau_1+t_2 k_y\left(k_z^2-k_x^2\right) \tau_2 \\
+\tau_3\left[\left(\lambda_z k_x+\lambda_z^{\prime} k_x k_y^2+\lambda_z^{\prime \prime}\left(k_x^3-3 k_x k_z^2\right)\right) \sigma_x\right. \\
+\left(\lambda_y k_y+\lambda_y^{\prime} k_y^3\right) \sigma_y \\
\left.+\left(\lambda_z k_z+\lambda_z^{\prime} k_z k_y^2+\lambda_z^{\prime \prime}\left(k_z^3-3 k_z k_x^2\right)\right) \sigma_z\right],
\end{gathered}
\label{Ham1}
\end{equation}
where the origin for the $k_i$ is the X(0,1,0) point (the formulas for the X$_1$(1,0,0) and X$_3$(0,0,1) points can be obtained by appropriate permutation of the subscripts of $k_i$ and $\sigma_i$), $\varepsilon_{0, \mathbf{k}}=\frac{\hbar^2}{2 m}\left(k_x^2+k_z^2+\beta k_y^2 \right)$, the $\lambda_i$ are SOC terms, and the $t_i$ are hopping terms. Eq.~(\ref{Ham1}) contains all operators $\tau_i\sigma_j$ allowed by symmetry, and we include terms up to order $k^3$. It can be diagonalized to yield eigenvalues
\begin{equation}
E_\pm(\mathbf{k})=\varepsilon_{0, \mathbf{k}}\pm\tilde{\varepsilon}_\mathbf{k},
\end{equation} where
\begin{equation}
\begin{array}{ll}
\tilde{\varepsilon}_\mathbf{k}= &
[ \ \lambda_z^2\left(k_x^2+k_z^2\right)+\lambda_y^2 k_y^2+t_1^2 k_x^2 k_z^2 \\
 & \ + 2 \lambda_y \lambda_y^{\prime} k_y^4 
  + \, 2\lambda_z \lambda_z^{\prime}\left(k_x^2+k_z^2\right) k_y^2 \\
 & \ + 2 \lambda_z \lambda_z^{\prime \prime}\left(k_x^4+k_z^4-6 k_x^2 k_z^2\right)
\ ]^{\frac{1}{2}},
\end{array}
\label{3rd_dispersion}
\end{equation}
keeping all terms up to $k^4$ under the square root in $\tilde{\varepsilon}_\mathbf{k}$. (The $t_2$ term is neglected since that enters as $k^6$.) Keeping terms related to order $k^3$ in spin-orbit ($\lambda_y \lambda_y^{\prime}$, $\lambda_z \lambda_z^{\prime}$, $\lambda_z \lambda_z^{\prime \prime}$) is required both formally and to describe the VHS and Fermi surfaces seen in our DFT calculations. To gain analytic insight, note that when the $\lambda_i=0$ (no SOC), $\tilde{\varepsilon}_\mathbf{k}=0$ for the lines $k_x=k_y=0$ and $k_y=k_z=0$. These are nodal lines protected by  non-symmorphic symmetries $(T \tilde{M}_{2, \hat{\bm{n}}})^2=-1$ where $T$ is time-reversal symmetry and $\tilde{M}_{2, \hat{\bm{n}}}$ is a glide mirror symmetry. Specifically, in $T \tilde{M}_{2,\hat{\bm{x}}}$ the $k_x$ nodal line is protected by $\tilde{M}_{2, \hat{\bm{x}}}=\{M_z|1/4,/14,0\}$. Along these nodal lines, SOC is the only interaction that splits this degeneracy, leading to anomalous pseudospin. Specifically, for the $k_z$ ($k_x$) nodal line, only $\sigma_z$ ($\sigma_x$) appears. Furthermore, using a symmetry analysis similar to that in Ref.~\cite{suh2023superconductivity}, it is possible to show that two pseudospin states related by $TI$ symmetry (where $I$ is inversion) can only couple to Zeeman fields along the $\hat{z}$ ($\hat{x}$) direction.

For an isotropic $s$-wave superconductor, which we assume to be the case for Ti$_4$Ir$_2$O, the SOC-driven Pauli limiting fields are determined by~\cite{suh2023superconductivity,Cavanagh:2023} 
\begin{equation}
\ln \frac{h_{P, \mathbf{\hat{h}}}}{h_0}=\langle-\ln \tilde{g}_{\mathbf{k}, \mathbf{\hat{h}}}\rangle_{\mathbf{k}},
\label{equ_hP}
\end{equation}
where $\mathbf{\hat{h}}$ is the direction of the applied magnetic field, $h_0$ is the usual Pauli limiting field in the absence of SOC, $\langle\ldots\rangle_{\mathbf{k}}$ denotes an average over the Fermi surface weighted by the momentum-dependent DOS, and $\tilde{g}_{\mathbf{k}, \mathbf{\hat{h}}}$ is the effective $g$-factor for a field oriented along $\mathbf{\hat{h}}$ and is given by 
\begin{equation}
\tilde{g}_{\mathbf{k}, \mathbf{\hat{h}}} =\sqrt{\frac{t_{1, \mathbf{k}}^2+t_{2, \mathbf{k}}^2+(\boldsymbol{\lambda}_{\mathbf{k}} \cdot \mathbf{\hat{h}})^2}{t_{1, \mathbf{k}}^2+t_{2, \mathbf{k}}^2+\boldsymbol{\lambda}_{\mathbf{k}}^2}}.
\label{g}
\end{equation}

Keeping only order $k$ terms in the $kp$ dispersion implies ellipsoidal Fermi surfaces around each X point. These Fermi surfaces are defined  by $\mu=\sqrt{\lambda_z^2 (k_x^2+k_z^2)+\lambda_y^2k_y^2}$, where $\mu$ is the chemical potential. It is possible to analytically show that  $h_{P, \hat{\mathbf{h}}} / h_0 = e \approx 2.718$ for all field orientations. This enhancement of the critical field is a consequence of the spin texture on the three ellipsoidal Fermi surfaces. Specifically, the spin quantization axis is along the surface normal to the ellipsoidal Fermi surfaces, implying that generically $|\tilde{g}_{\mathbf{k}, \mathbf{\hat{h}}}|<1$ (except for isolated points on the Fermi surface). 

Keeping up to linear terms in SOC and quadratic terms in the $kp$ dispersion results in two different scenarios depending on the sign of $\beta$ in $\varepsilon_{0, \mathbf{k}}=\frac{\hbar^2}{2 m}\left(k_x^2+k_z^2+\beta k_y^2 \right)$. 

\begin{figure} 
\centering
\includegraphics[width=0.5\textwidth]{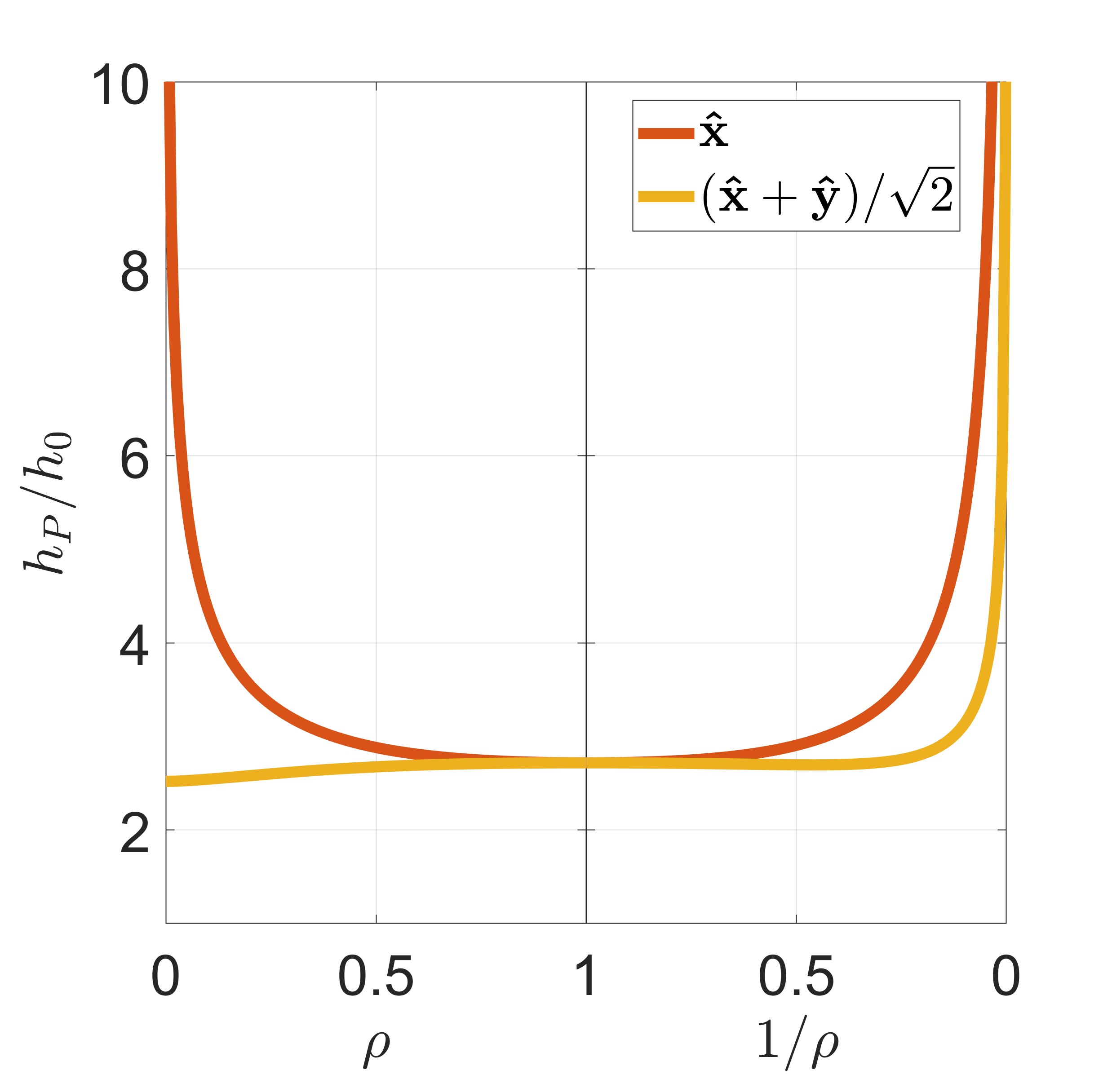}
\caption{The enhancement of the Pauli field for ellipsoidal Fermi surfaces near the X points. The ratio of the SOC-driven to the usual Pauli limiting field is plotted as a function of $\rho \equiv \frac{\lambda_y}{\lambda_z\gamma}$ on the left and of $1 / \rho$ on the right. These results are based on analytic formulas given in Supplemental Material~\cite{sm}. The enhancement is shown for fields along the $\mathbf{\hat{x}}$ and $\frac{\mathbf{\hat{x}}+\mathbf{\hat{y}}}{\sqrt{2}}$ directions.}
\label{fig3_hP_2nd_ana}
\end{figure}

When $\beta=\gamma^2>0$, the X point Fermi surfaces are ellipsoidal as was the case discussed above, and again an analytical result can be found for the critical field (see Supplemental Material~\cite{sm} for the analytic formulas). In this case, while the critical fields are enhanced (in some cases, the critical field can be formally divergent), the critical field is no longer isotropic as it was in the linear $k$ theory. Specifically, we find that the critical field for the field along the $(1,0,0)$ is always greater than or equal to that along the $(1,1,0)$ direction, as shown in Fig.~\ref{fig3_hP_2nd_ana}.

When $\beta=-\gamma^2<0$, a VHS appears in the dispersion.  It can be analytically shown that this VHS is due to a circle of saddle points defined by $k_x^2+k_z^2=\frac{m^2 \lambda_z^2}{\hbar^4}$ and $k_y=0$, which leads to a logarithmic divergence in the DOS.  In this case, numerical calculations reveal enhanced critical fields, typically by a factor of 2-3,  for the chemical potential in the vicinity of the VHS. When the chemical potential is at the VHS, both the critical field and the DOS are maximal since $\tilde{g}_{\mathbf{k}, \mathbf{\hat{h}}}$ vanishes at points on the circle of saddle points that form the VHS (see Supplemental Material~\cite{sm} for the circular Lifshitz transition, the DOS peak, and the $h_{P, \hat{\mathbf{x}}} / h_0$ peak).

Keeping the $k^3$ terms in SOC, as in Eq.~(\ref{3rd_dispersion}), yields a VHS that agrees with that found in our DFT calculations. Specifically, taking $\beta<0$, the product $\lambda_z\lambda_z'<0$, and $2 \lambda_z \lambda_z^{\prime} m^2 / \hbar^4+\gamma^2+\tilde{\gamma}^2<0$, where $\tilde{\gamma} \equiv \lambda_y / \lambda_z$, we find that the VHS discussed in the previous paragraph is shifted to $k_y\ne 0$ and no longer has a divergent DOS. Our DFT results confirm that $\lambda_z'$ is negative ($\lambda_z>0$) with a large magnitude, consistent with the conditions outlined above.

\begin{figure} 
\centering    
\includegraphics[width = 0.5\textwidth]{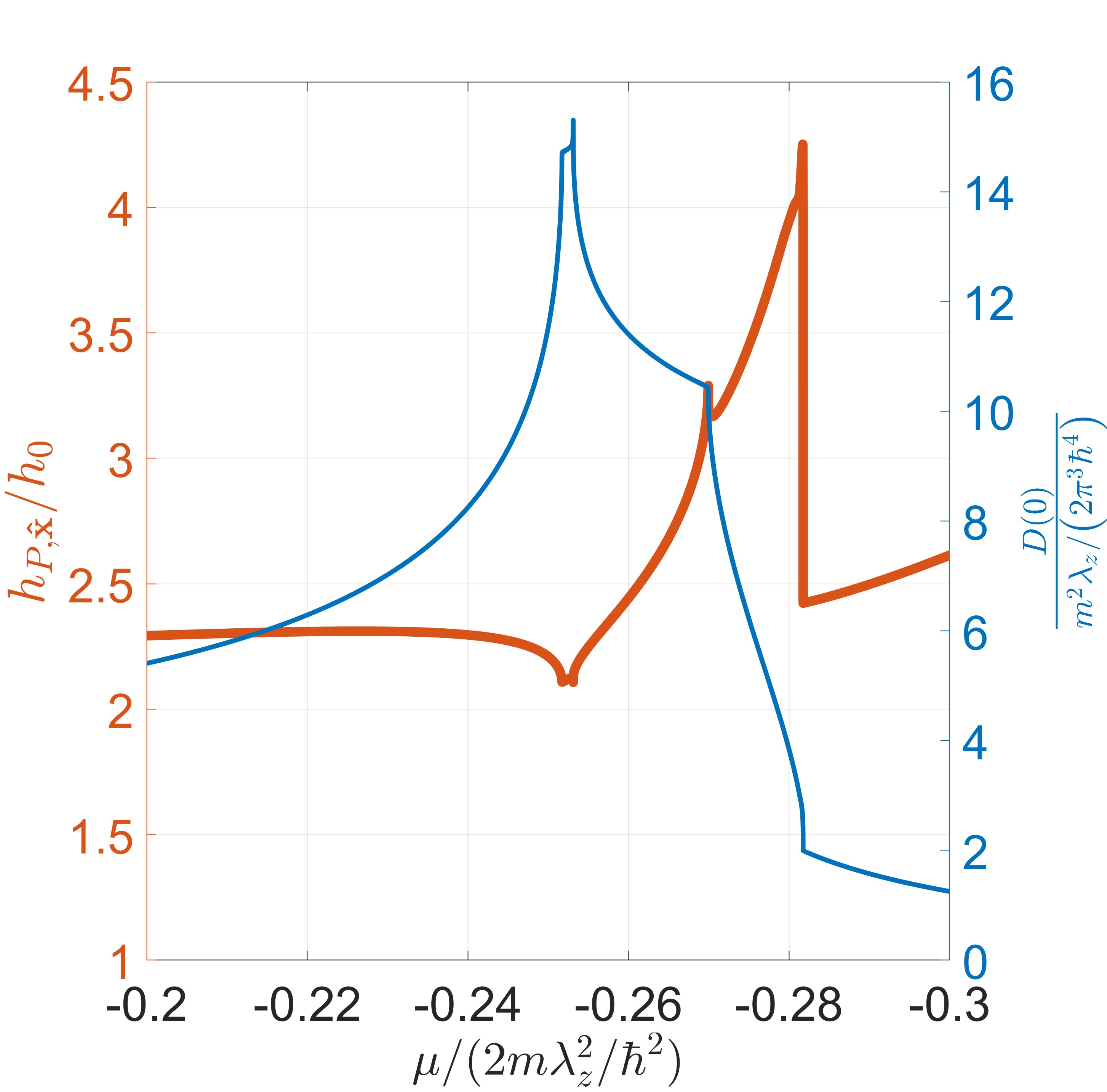}
\caption{The enhancement of the Pauli field and the DOS for $k^3$ $kp$ theory discussed in the text. The enhancement $h_{P, \hat{\mathbf{x}}} / h_0$ generally exceeds 2. The two peaks in $h_{P, \hat{\mathbf{x}}} / h_0$ from this theory arise because the region with a vanishing $g$-factor contains the area where the Lifshitz transition takes place.} 
\label{fig4_hP_3rd4} 
\end{figure}

\begin{figure*}
\centering
\includegraphics[width = 1\textwidth]{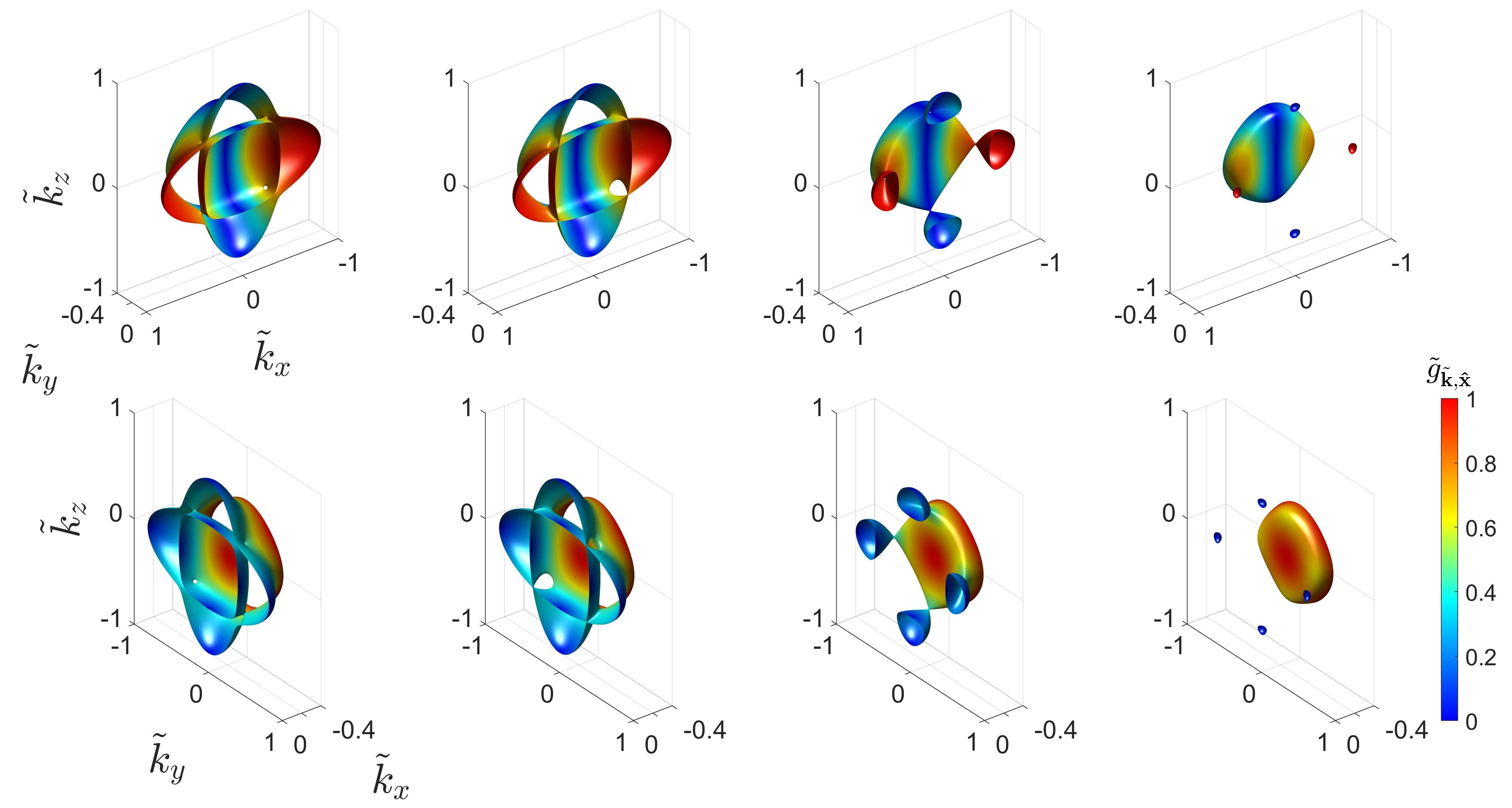}
\caption{Lifshitz transitions and reduced $g$-factors with changing the chemical potential. These are shown for momenta near the X(0,1,0) point (top) and the X$_1$(1,0,0) point (bottom). The panel for the X$_3$(0,0,1) point can be obtained by exchanging the labels of $\tilde{k}_z$ and $\tilde{k}_y$ in the X(0,1,0) panel. The applied field is along the $x$-direction. The color distribution reflects the reduced values of $\tilde{g}_{\tilde{\mathbf{k}}, \mathbf{\hat{x}}}$, which indicate the strength of the enhancement of the Pauli field. Notably, two of the six momentum lines exhibiting anomalous pseudospin are guaranteed by symmetry to have vanishing $g$-factors for this field orientation. This is shown as the `blue belts' in the X(0,1,0) panel, and they also appear in the symmetry-related X$_3$(0,0,1) panel (not displayed). These two lines have a large contribution to the enhancement of the Pauli field. (See Supplemental Material~\cite{sm} for details on the parameters and dimensionless quantities.)}
\label{fig5_FS_3rd4}
\end{figure*}

Choosing parameters that reveal the VHS Fermi surface, we numerically calculate the enhancement of the Pauli field $h_{P, \hat{\mathbf{x}}} / h_0$, as shown in Fig.~\ref{fig4_hP_3rd4}.
We observe that the enhancement $h_{P, \hat{\mathbf{x}}} / h_0$ generally exceeds 2. The underlying reason for this enhancement can be understood by examining the color distribution of  $\tilde{g}_{\tilde{\mathbf{k}}, \hat{\mathbf{x}}}$ in Fig.~\ref{fig5_FS_3rd4}. The reduced value of the effective $g$-factor suggests a significant enhancement of the Pauli field. Given the field in the $x$-direction, symmetry ensures that two of the six momentum lines exhibiting anomalous pseudospin have vanishing $g$-factors. These two lines are (0,1,u) on the X(0,1,0) face and (0,u,1) on the X$_{3}$(0,0,1) face, and they lead to the `blue belts', as shown in Fig.~\ref{fig5_FS_3rd4}. If we now consider the X(0,1,0) face, this contains two perpendicular anomalous pseudospin lines: one along (u,1,0) and another along (0,1,u). The anomalous pseudospin on (u,1,0) is oriented along the $x$-direction, while the anomalous pseudospin on (0,1,u) aligns with the $z$-direction, which is perpendicular to the applied $x$-direction field. This explains why only one `blue belt' appears on this face. On the anomalous pseudospin  lines, we have
$t_{1, \mathbf{k}}=0$ and $t_{2, \mathbf{k}}=0$. This implies that for a $x$-direction field, $\boldsymbol{\lambda}_{\mathbf{k}} \cdot \mathbf{\hat{h}}$ is solely determined by the values of $k_x$ near each X point. The absence of `blue belts' in the X$_{1}$(1,0,0) face is due to the Fermi surface having a finite $k_x$. However, for the X(0,1,0) and the X$_{3}$(0,0,1) faces, $k_x$ is zero along two lines of anomalous pseudospin, leading to a vanishing $g$-factor and an enhanced Pauli field.

{\it Discussion and conclusions.} 
Our analysis emphasizes anomalous pseudospin states near a VHS. However, for the chemical potential fixed to the VHS, there are also states, accounting for 35\% DOS away from the VHS (this percentage becomes larger as the chemical potential is tuned away from the VHS). For these states, the effective $g$-factor of the Pauli limiting field is not expected to be reduced (a reduced $g$-factor leads to an enhanced Pauli field), that is $g\approx 1$. The interplay between the reduced $g$-factor electrons near the VHS and the normal $g$-factor electrons away from the VHS can account for a variety of other observed properties in Ti$_4$Ir$_2$O. One such property is the pressure dependence of the upper critical field~\cite{shi2023pressure}. In particular, if increasing the applied pressure causes the states with a reduced $g$-factor to move away from the chemical potential, then the upper critical field will decrease as observed. Furthermore, our mechanism may more generally account for the varying Pauli limiting fields seen in $\eta$-carbide-type superconductors.  In particular, we can understand why some materials have critical fields that are comparable to the Pauli limiting field and others have enhanced critical fields. This depends on where the chemical potential sits relative to any VHS with anomalous pseudospin.

In addition, it has recently been observed that for fields on the order of the Pauli limiting field, specific heat measurements observe anomalies inside the superconducting state~\cite{hu2023thermodynamic}.  These anomalies have been taken as evidence for a Fulde-Ferrell-Larkin-Ovchinnikov (FFLO) state~\cite{Fulde:1964, Larkin:1965}. However, this behavior is a natural consequence of having both electrons with a reduced $g$-factor and electrons with a normal $g$-factor.  More specifically, it has been shown that in a two-band superconductor~\cite{barzykin2009magnetic, barzykin2007gapless} in which both bands have the same $g$-factor, but the two gaps are different, a robust first-order phase transition occurs when the smaller gap is suppressed by an external magnetic field. The system then enters a partially gapless state. Similar physics is expected if the two gaps are comparable but the $g$-factors of the two bands differ, as we suggest here. This implies that the electronic states with $g\approx 1$ will be much more strongly suppressed by the magnetic field than those with a reduced $g$-factor, resulting in a field-induced momentum-dependent gap suppression.

In summary, we argue that the surprisingly strong violation of the Pauli limit in the cubic superconductor Ti$_{4}$Ir$_2$O originates from two key factors. The first is a symmetry-required strong SOC and reduced $g$-factor for the electronic states near the X points; the other is a Van Hove singularity near the X points associated with a DOS peak. This explanation also accounts for the pressure dependence of the upper critical field and the observed high-field specific heat anomalies. Finally, our theory leads to two predictions: the first is an anisotropy in the paramagnetic response and the second is a field-driven momentum-dependent gap suppression.

\vglue 0.5 cm
{\it Acknowledgments.}  We thank Philip Brydon and Yue Yu for useful directions. DFA and HW were supported by the U.S. Department of Energy, Office of Basic Energy Sciences, Division of Materials Sciences and Engineering under Award No. DE-SC0021971 for $kp$ theory calculations. MW and TS were supported by National Science Foundation Grant No. DMREF 2323857 for DFT calculations. The Fermi surfaces in Fig.~\ref{FS} are visualized using FermiSurfer~\cite{kawamura2019fermisurfer}.

\bibliography{ref}

\end{document}


\title{Supplemental Material for \\ Large critical fields in superconducting Ti$_{4}$Ir$_2$O from spin-orbit coupling}
\author{Hao Wu}
\author{Tatsuya Shishidou} 
\author{Michael Weinert} 
\author{Daniel F. Agterberg}
\affiliation{Department of Physics, University of Wisconsin--Milwaukee, Milwaukee, Wisconsin 53201, USA}
\date{\today}

\maketitle

\section{Parameters from DFT}
\begin{figure}[th]
\begin{center}
	\includegraphics[width=0.9\columnwidth]{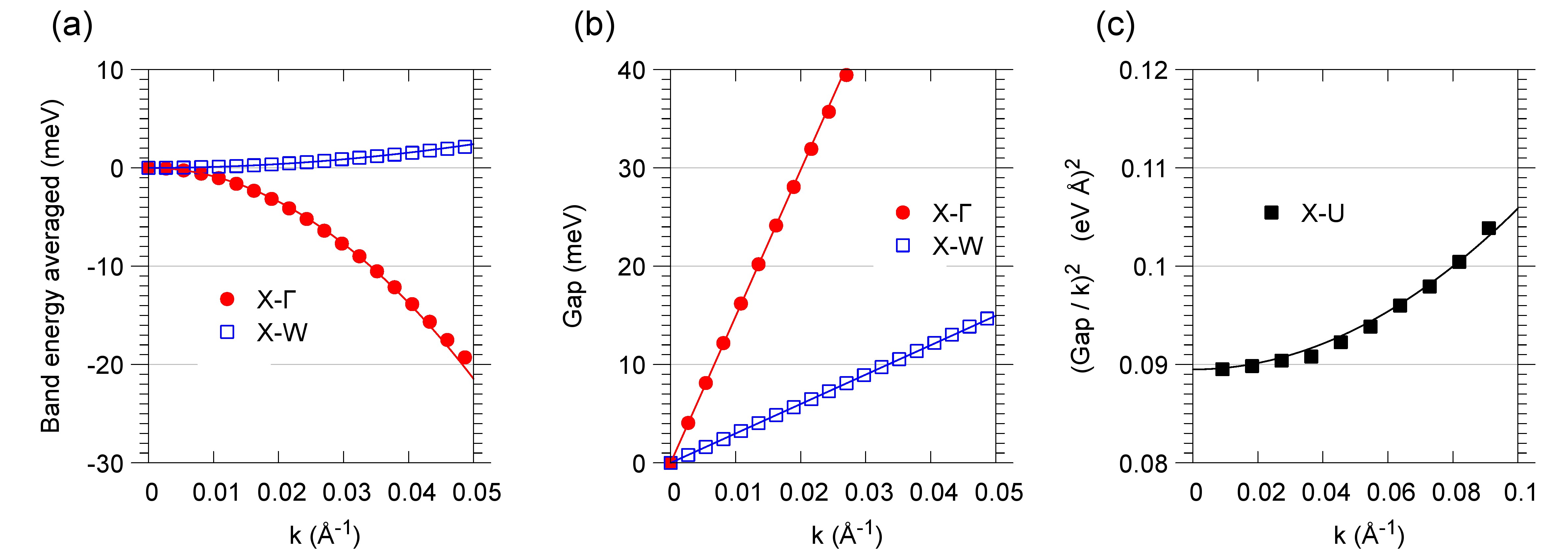}
\end{center}
\vspace{-3mm}
	\caption{Analysis for the DFT four bands (\#161-164) with spin-orbit coupling included.  Different $k$ directions from X point (X-$\Gamma$, X-W, and X-U) are examined.  (a) Averaged band energy to deduce the band curvature.  (b) Band gap whose linear slope gives the $k$-linear spin-orbit parameter.  (c) The square of the band gap divided by $k$ is shown along the X-U direction, which equals to $4 \left(\lambda_z^2+ (t_1/2)^2 k^2 \right)$.  In each panel, the estimated analytic curves are shown by solid lines.}
	\label{Fig:DFTparam}
\end{figure}

Our $kp$ Hamiltonian that describes the behavior of anomalous pseudospin around the X point is characterized by several parameters.  By referring to the DFT calculation, we can have an idea of their reasonable magnitudes.  From the analysis given in Fig.~\ref{Fig:DFTparam}, we evaluated $\left( \frac{\hbar^2}{2m}, \frac{\hbar^2}{2m}\gamma^2, t_1 \right) = (0.97, 8.58, 1.30 ) $ eV\,\AA$^2$ for $k$-quadratic terms;  $\left(\lambda_y, \lambda_z\right) = (0.75, 0.15 )$ eV\,\AA\ for $k$-linear spin-orbit.   For the third-order parameter, $t_2$, the magnitude seems too small to make an estimate. 

\section{Parameters for Fig. 5 in the main text}
We define the following dimensionless quantities to plot the Fermi surfaces: $\tilde{k}_{x(z)} \equiv \frac{k_{x(z)}}{2 m \lambda_z / \hbar^2}=\frac{k_{x(z)}}{0.155 \, \text{\AA}^{-1}}$, $\tilde{k}_y \equiv \frac{k_y}{2 m \lambda_z /\left(\hbar^2 \tilde{\gamma}\right)}$, $\tilde{\gamma} \equiv \frac{\lambda_y}{\lambda_z}$, $\tilde{\mu} \equiv \frac{\mu}{2 m \lambda_z^2 / \hbar^2}=\frac{\mu}{23.2 \, \mathrm{meV}}$, $\tilde{t}_1 \equiv \frac{t_1}{\hbar^2 / 2 m}$, $\tilde{\lambda}_z^{\prime} \equiv \frac{\lambda_z^{\prime}}{\left(\hbar^2 / 2 m\right)^2 \tilde{\gamma}^2 / \lambda_z}$, $\tilde{\lambda}_z^{\prime \prime} \equiv \frac{\lambda_z^{\prime \prime}}{\left(\hbar^2 / 2 m\right)^2 / \lambda_z}$, and $\tilde{\lambda}_y^{\prime} \equiv \frac{\lambda_y^{\prime}}{\left(\hbar^2 / 2 m\right)^2 \tilde{\gamma}^3 / \lambda_z}$.
We adopt the following DFT parameters: $\hbar^2 / 2 m = 0.97 \, \text{eV} \, \text{\AA}^2$, $\gamma = 2.97$, $t_1=1.30 \, \text{eV} \, \text{\AA}^2$, $\lambda_y=0.75$ eV$\text{\AA}$, and $\lambda_z=0.15$ eV$\text{\AA}$. For Fig. 5 VHS, we set $\tilde{\lambda}_z^{\prime}=-3, \, \tilde{\lambda}_z^{\prime \prime}=0.2$, and $\tilde{\lambda}_y^{\prime}=-0.04$.

\section{Analytic formulas when keeping up to linear terms in SOC and quadratic terms in the $kp$ dispersion}

We keep up to linear terms in SOC and quadratic terms in the $kp$ dispersion with $\beta=\gamma^2>0$. An approximate ellipsoidal Fermi surface is determined by $\varepsilon_{0, \mathbf{k}}-\mu = 0$. The ratio $h_{P, \hat{\mathbf{x}}} / h_0$ can be analytically expressed as a function of $\rho \equiv \frac{\lambda_y}{\lambda_z\gamma}$ for fields along the $(1,0,0)$ and $(1,1,0)$ directions. These formulas are given by
\begin{equation}
\frac{h_{P,\mathbf{\hat{x}}}}{h_0}= \begin{cases}\rho^{\frac{2}{3}} f_1\left(\rho\right) & \text { for } 0<\rho<1 \\ \rho^{\frac{2}{3}} f_2\left(\rho\right) & \text { for } \rho>1,\end{cases}
\end{equation}
\begin{equation}
\frac{h_{P, \frac{\mathbf{\hat{x}}+\mathbf{\hat{y}}}{\sqrt{2}}}}{h_0}= \begin{cases}\left(\frac{2 \rho^3}{\rho^2+1}\right)^{\frac{1}{3}} f_1\left(\rho\right) & \text { for } 0<\rho<1 \\ \left(\frac{2 \rho^3}{\rho^2+1}\right)^{\frac{1}{3}} f_2\left(\rho\right) & \text { for } \rho>1,\end{cases}
\end{equation}
where $f_1(x) \equiv\left(\frac{1+\sqrt{1-x^2}}{1-\sqrt{1-x^2}}\right)^{\frac{1}{2 \sqrt{1-x^2}}}$ and $f_2(x) \equiv \exp \left(\frac{\tan ^{-1} \sqrt{x^2-1}}{\sqrt{x^2-1}}\right)$. 

\section{circular Van Hove singularity}
\begin{figure} [!b]
\centering
\includegraphics[width = 0.8\textwidth]{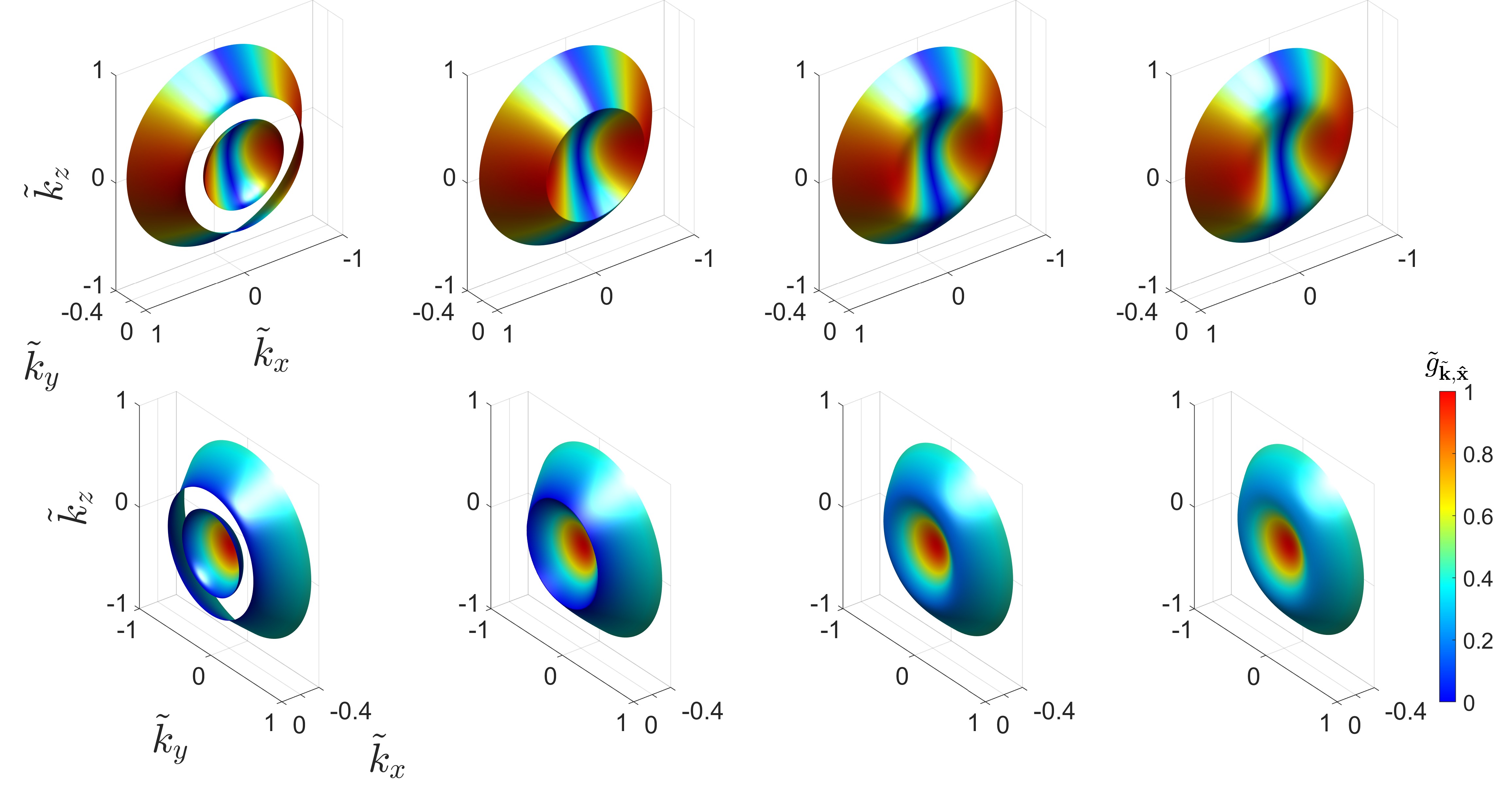}
\caption{Circular Van Hove singularity.  The parameters used are $\gamma = 2.97$ and $\tilde{\gamma} \equiv \frac{\lambda_y}{\lambda_z}=5$.}
\label{SM_FS_2nd}
\end{figure}

\begin{figure} [!b]
\centering
\includegraphics[width = 0.46\textwidth]{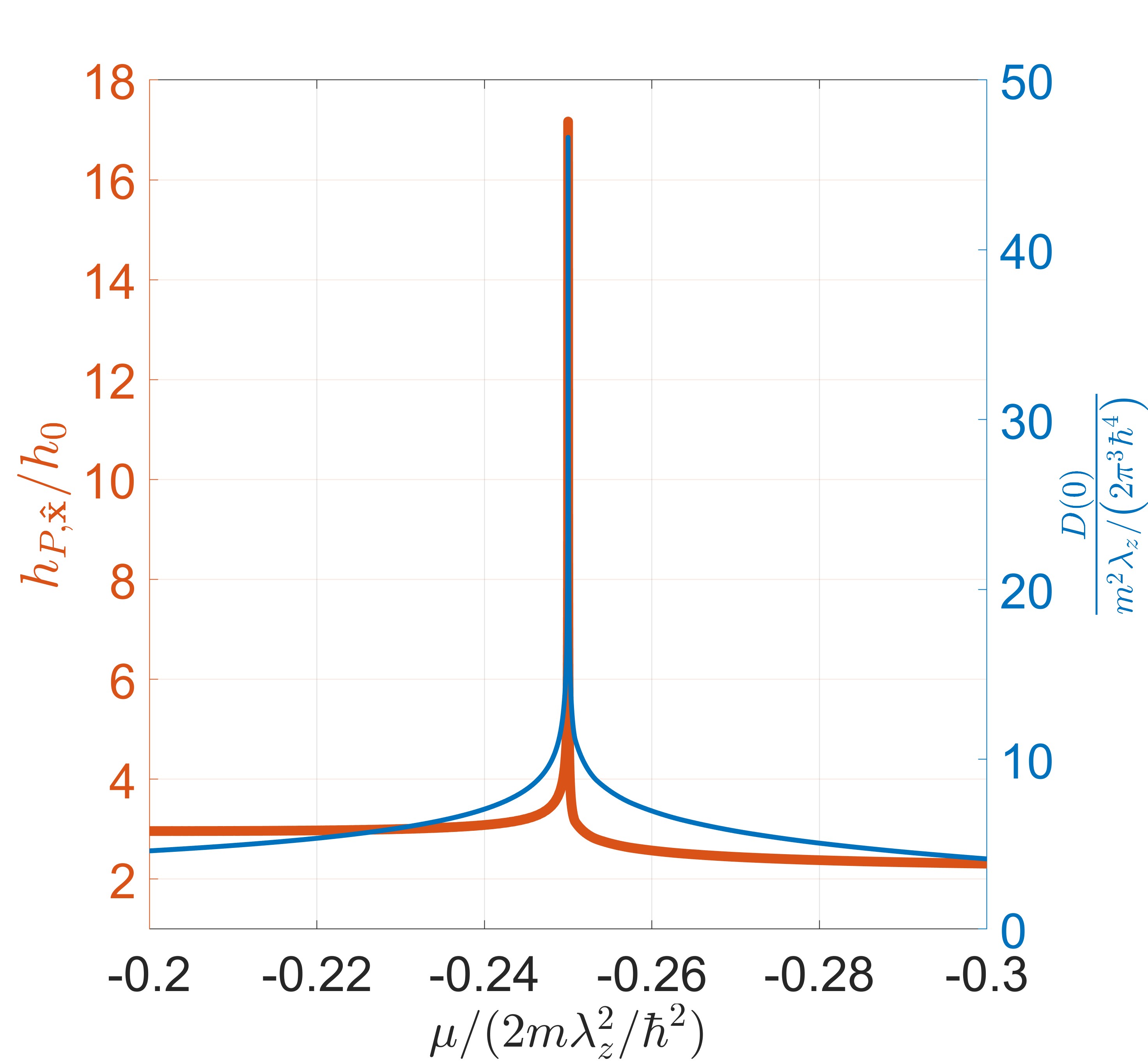}
\caption{Pauli field enhancement and density of states for circular Van Hove singularity scenario.}
\label{SM_hP_2nd}
\end{figure}

When we keep up to quadratic terms in the $kp$ dispersion with  $\beta=\gamma^2<0$, a circular Van Hove singularity appears, as illustrated in~\ref{SM_FS_2nd} for the Lifshitz transitions in the Fermi surfaces and in~\ref{SM_hP_2nd} for the Pauli field enhancement and the density of states peak. 

